\documentclass[12pt]{article}
\usepackage{amsmath,amssymb,amsthm,latexsym,graphicx,citesort}
\newtheorem{theorem}{Theorem}

\newtheorem{remark}[theorem]{Remark}

\newtheorem{definition}[theorem]{Definition}

\newcommand{\RR}{\mathbb{R}}
\newcommand{\ZZ}{\mathbb{Z}}

\newcommand{\G}{\Gamma}

\begin{document}

\title{Quantum graphs: an introduction and a brief survey\\
{\small This article is dedicated to the memory of Professor V. Geyler}}
\author{Peter Kuchment\\
Department of Mathematics\\
Texas A\& M University\\
College Station, TX, USA\\
kuchment@math.tamu.edu}
\maketitle
\begin{abstract}
The purpose of this text is to set up a few basic notions
concerning quantum graphs, to indicate some areas addressed in the quantum graph
research, and to provide some pointers to the literature. The pointers in many cases
are secondary, i.e. they refer to surveys in \cite{volume} or
elsewhere.
\end{abstract}
\section{Introduction}
We use the name ``quantum graph'' for a graph considered as a
one-dimensional singular variety and equipped with a differential (in some cases pseudo-differential)
operator
(``Hamiltonian''). There are manifold reasons for studying quantum
graphs. They naturally arise as simplified (due to reduced
dimension) models in mathematics, physics, chemistry, and
engineering (e.g., nanotechnology and microelectronics), when one considers propagation of
waves of various nature (electromagnetic, acoustic, etc.)
through a quasi-one-dimensional system (often a ``mesoscopic'' one)
that looks like a thin neighborhood of a graph. One can mention in particular
the free-electron theory of conjugated molecules in chemistry,
quantum wires, dynamical systems, photonic
crystals, thin waveguides, and many other applications.
We will provide the necessary references later on in this text.
The paper is intended to serve as a survey, a literature guide, and an introduction
that could be useful for reading other articles of this volume that are devoted to quantum graphs
and their applications.
One can find surveys and collections of papers on quantum graphs and related issues
in \cite{Exner_book,GnuSmi,Hurt,Keating,KoS1,KoS2,BCFK,WRM,WPRM,Ku_graphs1,Pokornyi,Ku02,KuISAAC}.

\section{Graphs and metric graphs}
A {\bf graph} $\Gamma$ consists of a finite or countably infinite
set of vertices $V=\{v_i\}$ and a set $E=\{e_j\}$ of edges
connecting the vertices. Each edge $e$ can be identified with a
pair $(v_i,v_k)$ of vertices, its endpoints.
In most cases of interest, directions of the edges will be irrelevant,
although it is sometimes more convenient to have
them assigned arbitrarily\footnote{In studies devoted to quantum chaos on graphs, it is common to count each edge twice, i.e. with both directions.}. Loops and multiple edges are allowed.

We denote by $E_v$ the set of edges incident to the
vertex $v$ (i.e., containing $v$) and will always assume that the degree
(valence) $d_v=|E_v|$ of any vertex $v$ is finite and positive. Thus, vertices
with no incident edges are not allowed (it will be clear later that for the quantum
graph purposes such vertices are irrelevant).

We introduce now an additional structure that makes $\Gamma$ a topological and
metric, rather than purely combinatorial, object.
\begin{definition}\label{D:metr_graph} A graph $\Gamma$ is said to be
a {\bf metric graph}\footnote{Sometimes the notions of a {\bf weighted
graph} or {\bf $\RR$-graph} are used instead.}, if its each edge $e$ is assigned a
positive length $l_e \in (0,\infty ]$ (edges of
infinite length are allowed).
\end{definition}
An edge $e$ can be identified with a
finite or infinite segment $[0,l_e]$ of the real line. We will fix such an
identification for each edge, which introduces a coordinate $x_e$
along it\footnote{This also introduces a preferred direction on the edge. As we have mentioned before, in some cases it is convenient to introduce two copies of each edge, which are equipped with opposite directions and correspondingly reversed coordinates.}. When this cannot lead to confusion, the subscript $e$ will be omitted
and the coordinate will be denoted $x$. This defines natural topology on the graph, which makes $\Gamma$ a topological space ($1D$ simplicial complex). This space is the union of all edges, where the ends corresponding to the same vertex are identified.
Graph $\Gamma$ can also be equipped with a natural metric. If a
sequence of edges $\{e_j\}_{j=1}^M$ forms a path, its length can be defined
defined as $\sum l_j$. For two vertices $v$ and $w$, the distance
$\rho (v,w)$ is defined as the minimal path length between them.
It is also easy to define the natural distance $\rho (x,y)$ between two
points $x,\,y$ of the graph that are not necessarily vertices. We
leave this to the reader.

In the case of infinite graphs (i.e., graphs with infinitely many vertices), sometimes
the following additional condition of finiteness arises:\\
$\bullet$ {\bf Finite ball volume.} For any positive number $r$ and any vertex $v$
there is only a finite set of vertices $w$ at a distance less than $r$ from $v$.
In particular, the distance between any two distinct vertices is
positive, and there are no finite length paths of infinitely many
edges. This matters only for graphs with infinitely many edges and is
usually satisfied in applications.

In some cases (e.g., when studying fractals or infinite quantum trees),
this assumption is too restrictive and needs to be abandoned.

If infinite edges are present, in most cases the following condition is assumed:\\
$\bullet$ {\bf Infinite leads.} The ``infinite'' ends of infinite
edges are assumed to have degree one. Thus, the graph can be
thought of as a graph with finite length edges with one or more additional
infinite ``leads'' or ``ends'' going to infinity attached to some
vertices. This situation arises naturally for instance in
scattering theory.  These ``infinite'' vertices are usually not treated as
vertices, so one can just assume that each infinite edge is a ray with a single vertex.

The graph is not assumed to be embedded into an Euclidean space (or Riemannian manifold).
In some applications such a natural embedding does exist (e.g., in modeling quantum
wire circuits or photonic crystals, see further sections), and in such cases the
coordinate along an edge is usually the arc length. In some other
applications, there is no natural embedding.

\begin{figure}[ht]
\begin{center}
\includegraphics{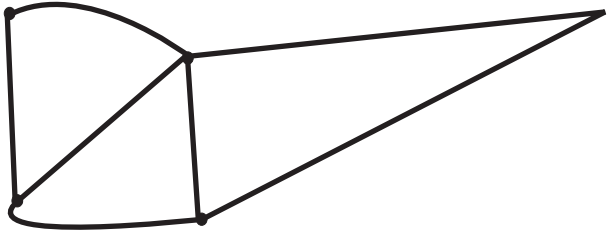}
\end{center}
\caption{A metric graph $\Gamma$.}\label{F:graph}
\end{figure}

It is useful to picture a metric graph $\Gamma$ as a one-dimensional
simplicial complex, each $1D$ simplex (edge) of which is equipped
with a smooth structure, with singularities arising at junctions
(vertices) (see Fig. \ref{F:graph}).

The points of a metric graph are not only its vertices, but all intermediate points
$x$ on the edges as well. When we say a ``function $f(x)$ on $\Gamma $,'' we mean
that the values $f(x)$
are defined along the edges, not just at the vertices, as is the case in discrete models.
Having the coordinate $x$, one can define the natural Lebesgue measure $dx$
on $\G$. The notions of measurability and integrability can be now introduced,
which enable one to define some standard function spaces on the graph:

\begin{definition}\label{D:spaces_l2}
 The space $L_2(\Gamma )$
on $\Gamma $ consists of functions that are measurable and square
integrable on each edge $e$ and such that
$$
\|f\|_{L_2(\Gamma)}^2=\sum\limits_{e \in E}\|f\|_{L_2(e)}^2 <
\infty.
$$
In other words, $L_2(\Gamma )$ is the orthogonal direct sum of
spaces $L_2(e)$.
\end{definition}
\noindent The finiteness of the sum condition is relevant for infinite graphs only.

Due to the presence of the coordinate $x$ along the edges, one can discuss
differentiability of a function $f$ on each edge (but not on the whole graph). This
leads to the following definition:
\begin{definition}\label{D:spaces_sob}
The Sobolev space $H^1(\Gamma)$
consists of all {\bf continuous} functions on $\Gamma$ that belong
to $H^1(e)$ for each edge $e$ and such that
$$
\sum\limits_{e \in E} \| f\|_{H^1(e)}^2 < \infty.
$$
\end{definition}
The continuity requirement in the definition of the Sobolev space $H^1$ means
that the functions on all edges adjacent to a vertex $v$ assume the same value at $v$.
This is a natural condition for one-dimensional $H^1$-functions. However,
there seems to be no natural definition of Sobolev spaces
$H^k(\Gamma)$ of order $k$ higher than $1$, due to the lack of natural conditions at vertices.
As we will see, these conditions might be different for different Hamiltonians (see details
later on in the following sections). Again, the finiteness of the sum condition is superfluous,
unless the graph is infinite.

An interesting moduli space of metric graphs\footnote{The name $\RR$-graphs was used there.} with a
fixed fundamental group was introduced twenty years ago in \cite{CullVogt} (see also
\cite{KrstVogtm,BridsVogt} and the survey \cite{Vogt}). It was used for studying outer automorphisms
of free groups (that is why the name ``Outer space'' is common). In particular, various natural
compactifications of this space have been introduced and studied.
Although this space has never been used in quantum graphs research, the author has a feeling that in
some problems its use might become beneficial.

\section{Quantum graphs}
In order to make a metric graph a quantum one, an additional structure is needed:
a differential (or sometimes more general) operator (Hamiltonian) on $\Gamma$, which is mostly, but not always, required to be self-adjoint. The most frequently studied operators of interest are acting as follows:
\newline the negative second derivative
\begin{equation}
f(x)\rightarrow -\frac{d^2f}{dx^2},  \label{E:deriv_2}
\end{equation}
a more general Schr\"{o}dinger operator

\begin{equation}
f(x)\rightarrow -\frac{d^2f}{dx^2}+V(x)f(x),  \label{E:electr}
\end{equation}
or a still more general magnetic Schr\"{o}dinger operator
\begin{equation}
f(x)\rightarrow \left( \frac 1i\frac d{dx}-A(x)\right)
^2f(x)+V(x)f(x). \label{E:electromagn}
\end{equation}
It is clear that the definition of
such an operator is not complete, till its domain is described.
For ``decent'' potentials $V$ and $A$, the natural requirement coming from the standard ODE theory is that $f$ belongs to the Sobolev
space $H^2(e)$ on each edge $e$. What is still missing, is having appropriate boundary value conditions at the vertices (\textbf{vertex conditions}). We will address these in the next section.

Possibilities of more general scalar or matrix differential or pseudo-differential operators
will be mentioned at a later stage. We will concentrate here on the most common scalar second order
differential operators, and for simplicity of exposition
on (\ref{E:deriv_2}).

\begin{definition}\label{D:qgraph}
A {\bf quantum graph} is a metric graph equipped with the operator
${\mathcal H}$ (Hamiltonian) that acts as the negative second order derivative along
edges\footnote{More general Hamiltonians also arise and will be discussed later in this text.} and is accompanied by ``appropriate'' vertex conditions.
\end{definition}

\section{Vertex conditions}

We will describe now what kind of boundary conditions one can add
to the differential expression (\ref{E:deriv_2}) in order to
create a ``reasonable'' operator. In most cases, being ``reasonable'' will mean being self-adjoint.

Standard Sobolev trace theorems say that a function $f\in H^2(e)$ and its first derivative have correctly defined values at the endpoints of the edge $e$. Already the second order derivatives of $H^2$ functions do not have traces at the vertices. It is thus clear that the vertex conditions may involve only the values of $f$
and $df/dx$ at a vertex $v$. In principle, the conditions can mix the values at different vertices (e.g., periodicity condition for a function $f$ on a segment $e$ does exactly that). We, however, will concentrate at the moment on the {\bf local vertex conditions} only, i.e. those that involve the values of functions and their derivatives at a single vertex at a time. We will see soon that the general, non-local, case can be reduced to the local one.

To avoid some technical details, we restrict ourselves here to {\bf finite graphs}.
I.e., we assume that the number of edges $|E|$ (and
hence the number of vertices $|V|$) is finite. The edges are
still allowed to have infinite length. One can find discussion of infinite graphs in \cite{Carlson5,EG,Ku_graphs2,Pankr,Ku_Vainb,Solom,SoSo}, as well as in Section \ref{S:inf}.

A typical vertex condition is what is often called the ``Kirchhoff''\footnote{This name, albeit often used, is not too appropriate.} condition:
\begin{equation}
\left\{
\begin{array}{l}
f(x)\text{ is continuous on }\Gamma \\
\text{and}\\ \text{at each vertex }v\text{ one has }\sum\limits_{e
\in E_v }\frac{df}{dx_e}(v)=0
\end{array}
\right..  \label{E:delta0_cond}
\end{equation}
Here the sum is taken over all edges $e$ incident to the vertex
$v$ and the {\bf derivatives are taken in the directions away
from the vertex} (we will call these ``outgoing directions'')\footnote{We will adhere
to this agreement about outgoing differentiations in all cases when these conditions are
involved.}. Sometimes (\ref{E:delta0_cond}) is called by the more appropriate name
{\bf Neumann condition} (to satisfy all the parties, we will use the conciliatory name {\bf Neumann-Kirchhoff conditions}). Indeed, at the ``loose ends'' (vertices of degree
1) it turns into the actual Neumann condition. Besides, as the
Neumann boundary condition for Laplace operator, it is natural.
I.e., as we will see later, the domain of the quadratic form of the
corresponding operator does not require any conditions on a function besides being in $H^1(\Gamma)$ (and thus continuous). Another useful remark is that under the boundary
conditions (\ref{E:delta0_cond}) one can eliminate all vertices of
degree 2, connecting the two adjacent edges into one smooth edge.

For local vertex conditions, it is sufficient to address the problem of
describing the conditions for a single junction of $d$ edges at a vertex
$v$ (a ``star graph''). Since along each edge our operator acts as a second order operator, one expects to establish two conditions per an
edge, and hence at each vertex the number of conditions must
coincide with the degree $d$ of the vertex. As we have already mentioned,
for functions in $H^2$ on each edge, the conditions may involve only the
boundary values of the function and its derivative. Then the most general
form of such (homogeneous) condition clearly is

\begin{equation}
A_vF(v)+B_vF'(v)=0.
\label{E:condit}
\end{equation}
Here $A_v$ and $B_v$ are $d \times d$ matrices, $F(v)$ is the column vector
$(f_1(v),...,f_d(v))^t$ of the values at the vertex $v$ that function $f$ attains along
all edges incident to $v$ (e.g., if $f$ is continuous, all these values will be the same), and $F'(v)=(f_1'(v),...,f_d'(v))^t$ is the column vector of the
values at $v$ of the derivatives taken along these edges in the
outgoing directions.
The rank of the $d \times 2d$ matrix $(A_v,B_v)$ must be equal to
$d$ (i.e., maximal) in order to ensure the correct number of
independent conditions.

One can describe completely all conditions
(\ref{E:condit}) that guarantee self-adjointness of the resulting operator.
This can be done by either using the von Neumann theory of extensions
of symmetric operators (as for instance described in
\cite{AkhGlaz}), or by its more recent version that amounts to
finding Lagrangian planes with respect to the complex symplectic
boundary form that corresponds to the maximal operator (see for
instance \cite{Everitt1,Everitt2,Everitt3,Har,N3,Pav2,Post}
for the accounts of this approach that goes back at least as far
as \cite{Naim}). The next theorem contains three different descriptions
of all self-adjoint vertex conditions. It combines the results from
\cite{KoS1,Har,FKW,Ku_graphs1}. Experience shows that all three of these descriptions are useful in various circumstances.

\begin{theorem}\label{T:BC} Let $\Gamma$ be a metric graph with finitely
many edges. Consider the operator ${\mathcal H}$ acting as
$-\dfrac{d^2}{dx_e^2}$ on each edge $e$, with the domain
consisting of functions that belong to $H^2(e)$ and certain local vertex conditions involving vertex values of functions and their derivatives. The operator is self-adjoint,
if and only if the vertex conditions can be written in one (and thus any)
of the following three forms:
\begin{description}
\item[A] Conditions (\ref{E:condit}) at each
vertex, where $\{ A_v,\,B_v\,|\,\, v \in V\}$ is a collection of
matrices of sizes $d_v \times d_v$ such that
\begin{itemize}
\item The $d_v\times 2d_v$ matrix
$(A_v\,B_v)$ has the maximal rank.
\item The matrix $A_vB_v^*$ is self-adjoint.
\end{itemize}

\item[B] For every vertex $v$ of degree $d_v$,
there exists a unitary $d_v\times d_v$ matrix $U_v$ such that the vertex conditions at $v$ are
\begin{equation}\label{E:cond_harmer}
    i(U_v-\mathbb{I})F(v)+(U_v+\mathbb{I})F^\prime(v)=0,
\end{equation}
where $\mathbb{I}$ is the $d_v\times d_v$ identity matrix.

\item[C] For every vertex $v$ of degree $d_v$,
there are two orthogonal (and mutually orthogonal) projectors
$P_v$, $Q_v$ operating in $\mathbb{C}^{d_v}$ and an invertible
self-adjoint operator $\Lambda_v$ operating in the subspace
$(1-P_v-Q_v)\mathbb{C}^{d_v}$ (either $P_v$, $Q_v$, or
$C_v := 1-P_v-Q_v$ might be zero), such that the functions $f$ in the
operator domain satisfy at each vertex $v$ the following boundary conditions:
\begin{equation}
\begin{cases}
P_vF(v)=0 \mbox{ - the ``Dirichlet part''}, \\
Q_vF'(v)=0 \mbox{ - the ``Neumann part''},\\
C_vF'(v) =\Lambda_v C_vF(v) \mbox{ - the ``Robin part''}.\\
\end{cases}\label{E:cond_kuch}
\end{equation}
\end{description}
\end{theorem}

\begin{remark}\indent

1. It is not hard to notice that the representation (\ref{E:condit}) of vertex conditions is not unique: multiplying matrices $A$ and $B$ from the left by any invertible matrix $C$ does not alter the conditions. On the other hand, (\ref{E:cond_harmer}) or (\ref{E:cond_kuch}), which clearly are particular cases of (\ref{E:condit}), parametrize conditions uniquely.

 2. Equivalence of (\ref{E:cond_harmer}) and (\ref{E:cond_kuch}) is rather straightforward. It is also not hard to show that (\ref{E:condit}) can be reduced to (\ref{E:cond_harmer}) or (\ref{E:cond_kuch}).
\end{remark}

Vertex conditions can be described also in a different manner, usually adopted in studying quantum chaos on graphs \cite{GnuSmi,KS,KS2}. It involves prescribing how waves scatter at each vertex. One can find discussion of relations between these descriptions and of the ones in the theorem above in the paper \cite{Berk} in \cite{volume}.

\subsection{Quadratic form}

It is easy to describe the quadratic form of the operator
${\mathcal H}$ corresponding to the (negative) second derivative along
each edge, with self-adjoint vertex conditions written in the form (C) of the preceding theorem.

\begin{theorem}\label{T:qform}
The quadratic form $h$ of ${\mathcal H}$ is given as
\begin{equation}\label{E:qform}
\begin{array}{c}
h[f,f]= \sum\limits_{e \in E} \int\limits_e |\frac{df}{dx}|^2dx +
\sum\limits_{v \in V}\langle \Lambda_vC_vF,C_vF\rangle,
\end{array}
\end{equation}
where $\langle , \rangle$ denotes the standard hermitian inner
product in $\Bbb{C}^{\dim C_v}$. The domain of this form consists of all
functions $f$ that belong to $H^1(e)$ on each edge $e$ and satisfy
at each vertex $v$ the condition $P_vF=0$.

Correspondingly, the sesqui-linear form of ${\mathcal H}$ is
\begin{equation}\label{E:sesqform}
h[f,g]= \sum\limits_{e \in E} \int\limits_e
\frac{df}{dx}\overline{\frac{dg}{dx}}dx + \sum\limits_{v \in
V}\langle \Lambda_vC_vF,C_vG\rangle.
\end{equation}
\end{theorem}

\subsection{Examples of boundary conditions}\label{S:BC_examples}

In this section we list some examples of vertex conditions that arise rather often.

\subsubsection{$\delta$-type conditions} These vertex conditions are defined as
follows:
\begin{equation}
\left\{
\begin{array}{l}
f(x)\text{ is continuous on }\Gamma \\
\text{and} \\
\text{at each vertex }v\,,\,\sum\limits_{e \in
E_v}\frac{df}{dx_e}(v)=\alpha _v f(v)
\end{array}
\right. .  \label{E:delta_cond}
\end{equation}
Here $\alpha_v $ are some fixed numbers. One can recognize these
conditions as an analog of conditions one obtains for the
Schr\"{o}dinger operator on the line with a $\delta$ potential,
which explains the name. The self-adjointness condition is satisfied if
and only if all numbers $\alpha_v$ are real. When $\alpha_v=0$, one arrives to the previously considered Neumann-Kirchhoff conditions.

\subsubsection{$\delta^\prime$-type conditions} These are similar to the $\delta$-type ones, with the roles of the vertex values of functions and their derivatives switched:
\begin{equation}
\left\{
\begin{array}{l}
\mbox{The values} \frac{df}{dx_e}(v) \mbox{ are independent of }e \mbox{ at any vertex }v.\\
\text{and} \\
\text{at each vertex }v\,,\,\sum\limits_{e \in
E_v}f_e(v)=\alpha _v \frac{df}{dx_e}(v)
\end{array}
\right. .  \label{E:deltaprime_cond}
\end{equation}
When $d_v=2$, these conditions correspond to the symmetrized version of what is usually called
$\delta^\prime$ conditions at a point. The true counterpart is provided in \cite{Ex6}.

\subsubsection{Decoupling conditions} There exist vertex conditions that essentially split
the graph into unrelated edges. One can consider for instance the {\bf vertex Dirichlet conditions},
that force the functions from the domain of the operator to vanish at all vertices. Then the operator
is the direct sum of the operators on each edge $e$ with Dirichlet conditions at the ends. Similar
thing happens if one enforces {\bf vertex Neumann conditions} (not to be mixed up with the
Neumann-Kirchhoff conditions), requesting that derivatives along each edge vanish at the vertices.
In both these cases, the topology of the graph is irrelevant from the quantum graph point of view.

\subsection{Non-local conditions and turning a quantum graph into a single ``rose''}

\begin{figure}[ht!]
\begin{center}
\includegraphics{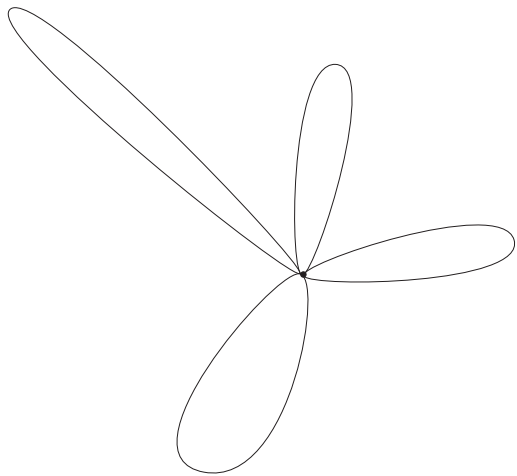}
\end{center}
\caption{Graph $\Gamma$ turned into a single ``rose.''}\label{F:rose}
\end{figure}

The above discussion of the decoupling conditions leads to the understanding that the whole topology
of the quantum graph is contained in the vertex conditions only. In particular, one can identify all
vertices of a quantum graph into one ``super-vertex'' $v_0$, so the graph becomes just a ``rose'' of
several petals (each edge bends into a loop), and all vertex conditions are written at this single
vertex (see Fig. \ref{F:rose}).
If one starts with some special type of vertex conditions, for instance Neumann-Kirchhoff ones,
this procedure will not preserve this special type. On the other hand, if one needs to work with
most general vertex conditions, then one can as well assume that the graph is just a rose of several
loops at a single vertex. This also shows that non-local conditions can be turned into local ones at
this single ``super-vertex'', and thus description of self-adjoint local conditions given in
Theorem \ref{T:BC} can be applied to non-local conditions as well.

\subsection{Non-selfadjoint conditions}

It is sometimes needed to consider more general Hamiltonians than the self-adjoint ones. For instance, one might be interested in accretive or dissipative Hamiltonians. The paper \cite{KoSsemigr} in \cite{volume} provides criteria under which the vertex conditions lead to such operators.

\subsection{Conditions involving spectral parameter} Sometimes vertex conditions arise that involve the spectral parameter $\lambda$. This usually happens when the ideal vertex corresponds to an object with some internal structure in a real world problem, see \cite{KZ2,KK02,ExPost,ExPost2,Ku02} for various examples.

\subsection{Realization of vertex conditions} It is an interesting, and not completely resolved
question of which of all possible vertex conditions can arise in practical problems leading to
quantum graph models. See, for instance, \cite{Ex5,Cheon,Exner_cond} for some partial results.

\section{Motivations for quantum graph models}

As it has already been mentioned, quantum graph models (often under different names, e.g. quantum
networks) come from various areas of mathematics, physics, and engineering, e.g. see the survey
\cite{Ku02} for details. Here we just provide some token (and incomplete) pointers to applications
in dynamical systems and probability theory \cite{Freidlin,Freidlin2,Freidlin3}, spectral theory of
differential operators on manifolds and in singular domains \cite{Colin,EH,evans2}, chemistry
(including studying carbon nanostructures)
\cite{Ex8,korotyaev:06,korotyaev:06_magn,amovilli:04,Gri1,Gri2,RiB,Schap,KuchPost,RuS},
superconductivity theory \cite{Al,deG,RS2,RS3,RS4}, photonic crystal theory
\cite{FK2,FK4,Ku01,KK99,KK02,Paiv,AKK}, microelectronics and  waveguide theory
\cite{ExS1,ExS2,ExKov,ExVug,Sirko,Krejcirik,ML,X,Pavl1,Pavl3,Pavl4,MolchVainb_slow},
biology \cite{Carlson,Nic,Carlson_lung}, acoustics \cite{acoustics}, quantum Hall effect
\cite{Buttiker}, and many others. Another reason for studying quantum graphs is that they
often offer a simplified, but still non-trivial models for complex phenomena, such for instance
as electron propagation in multiply connected media \cite{Avr,ARZ}, Anderson localization
\cite{AizSimsWarz1,AizSimsWarz2,AizSimsWarz3,AizSimsWarz4,AizSimsWarz5,HisPost,Exner_loc},
quantum chaos \cite{GnuSmi,Keating,KS,KS2,KS3}, and some problems of quantum field theory
\cite{Mintchev,Fulling,Fulling2,BHW,wilson,Fulling_vac,Fulling_vac2}. Quantum graph
(also called quantum network) models have also
been used for quite a while as toy models for quantum mechanics \cite{Montroll}.

\section{Justification of the quantum graph model for waves in narrow branching media}
\begin{figure}[ht!]
\begin{center}
\scalebox{0.4}{\includegraphics{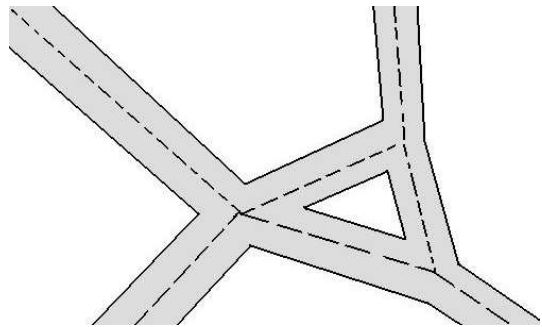}}
\end{center}
\caption{A narrow branching domain (shaded) with the approximating graph (dotted lines.)}\label{F:fat}
\end{figure}

One of the most important sources of quantum graphs is an attempt to model waves of various nature (acoustic, electromagnetic, electron, etc.) propagating in thin (often mesoscopic or nano-scale) branching media by waves propagating on graphs.
 Mathematically speaking, one deals with a partial differential operator (say, Laplace operator, or more general Schr\"{o}dinger operator with Dirichlet or Neumann boundary conditions) in a narrow branching domain that resembles a fattened graph (see Fig. \ref{F:fat}).
 Since studying such an operator is very hard both analytically and numerically, one wonders whether one can approximate properties (e.g., the spectrum) of the operator by those of an operator on a graph itself. This turns out to
be a highly non-trivial question, which has attracted attention of many researchers. Survey
\cite{Gri_survey} in \cite{volume} provides a nice overview of the exciting recent developments in this
problem and of the mathematics involved, as well as comprehensive references.

\section{Spectral properties}

Among the properties of quantum graph Hamiltonians that have attracted most attention are those related to their spectra: the types of spectra that can arise, spectral gaps, spectral asymptotics and statistics, regular and generalized eigenfunctions, scattering theory, etc. In this section, we will glance over the various topics that have been considered.

\subsection{Finite graphs}

In the case of a compact graph (i.e., a finite graph with all edges of a finite length), standard
Sobolev embedding theorems imply discreteness of the spectrum (see, e.g., \cite{Ku_graphs1} for this
trivial folklore result). One of the main achievements have been explicit trace formulas that can
be derived for the quantum graph case and which are the cornerstone for many further developments,
e.g. inverse problems and quantum chaos studies. See, for instance,
\cite{Roth,GnuSmi,KS,KS2,KS3,Winn} and the survey \cite{Bolte} in \cite{volume}.

Another large area of research is ``quantum chaology'' on quantum graphs, which deals with the
spectral statistics of quantum graph operators. It is surveyed in \cite{GnuSmi}, as well as in
\cite{Keating} in \cite{volume}.

Generic and extremal properties of quantum graph spectra, e.g. simplicity were studied in
\cite{Fr_graph,Fr_graph2}.

Determinants of quantum graph Hamiltonians are considered in \cite{ACD,Fr_graph_det}.

Index theorems for quantum graphs in relation to heat kernel asymptotics were considered in
\cite{FKW}.

\subsection{Finite graphs with infinite leads attached}

Assume now that several infinite leads are attached to a compact graph. One can think of this
situation as of a simple ``star graph'', where an ``obstacle'' (a compact graph) is inserted
into the vertex. Then the natural questions to study are about limiting absorption and scattering.
These have been addressed in quite a few publications, e.g. in \cite{Ong,G,GP,KoS3,MP,N1,N2,N3,PF}.
One of the simple but crucial ideas in this topic in particular, and in quantum (as well as
combinatorial) graph research in general, is usage of the so called Dirichlet-to-Neumann map that
enables one to eliminate the scatterer sitting at the vertex and replace it with an energy dependent
vertex condition, see for instance \cite{Ku_graphs2,Ong,Fox,Pavl1} for the quantum and \cite{GrigNekr}
for the combinatorial graph case. This trick, also well known in matrix theory as Schur complement
and in physics as Feshbach formula, is particularly useful when treating self-similar (e.g., fractal)
structures. It is also responsible for the ``decoration'' mechanism for spectral gap opening,
discussed below.

\subsection{Infinite graphs}\label{S:inf}
The case of an infinite quantum graph (i.e., a graph with infinitely many edges), even under the
finite ball volume condition, is rather complex and not much can be said about it in general. For
instance, all kinds of spectra can arise: pure point, singular continuous, and absolutely continuous.
There are only a few general things that can be established. First of all, due to the presence of
continuous spectrum, one cannot test whether a given point $\lambda$ belongs to the spectrum by
checking existence of a corresponding eigenfunction. However, there exist two types of PDE theorems
that help with this difficulty, allowing one to test the spectrum by looking for {\bf generalized
eigenfunctions}, i.e. those that do not decay fast enough (or do not decay at all) to be  true
eigenfunctions.

The first type are the generalized eigenfunction expansions (see \cite{Berez,BerezShub} and
references therein) which say that under some conditions on a self-adjoint partial differential
operator in $\RR^n$, there exists a set of values of $\lambda$ of the full spectral measure and for
each point of this set a generalized eigenfunction of a controlled (usually polynomial) growth,
which form a complete family in $L^2$. One can find a general framework of this kind of theorems
nicely described in an appendix to \cite{BerezShub}. This general technique can be, and has been
applied to quantum graphs \cite{HisPost}.

In a converse direction, Schnol' type theorems \cite{Schnol,Glaz,Cycon} claim that existence for some
$\lambda$ of a sub-exponentially growing generalized eigenfunction implies that $\lambda$ is in the
spectrum. A Schnol' type theorem holds also for infinite graphs under some mild conditions
\cite{Ku_graphs2}. Notice that Schnol' theorem needs to be transferred from PDEs in $\RR^n$ to
infinite graphs with some modification \cite{Ku_graphs2}. Direct transfer would lead to the
requirement that the volume of the ball of radius $r$ grows sub-exponentially with the radius,
which would exclude the case of trees.

\begin{figure}[ht!]
\begin{center}
\scalebox{0.3}{\includegraphics{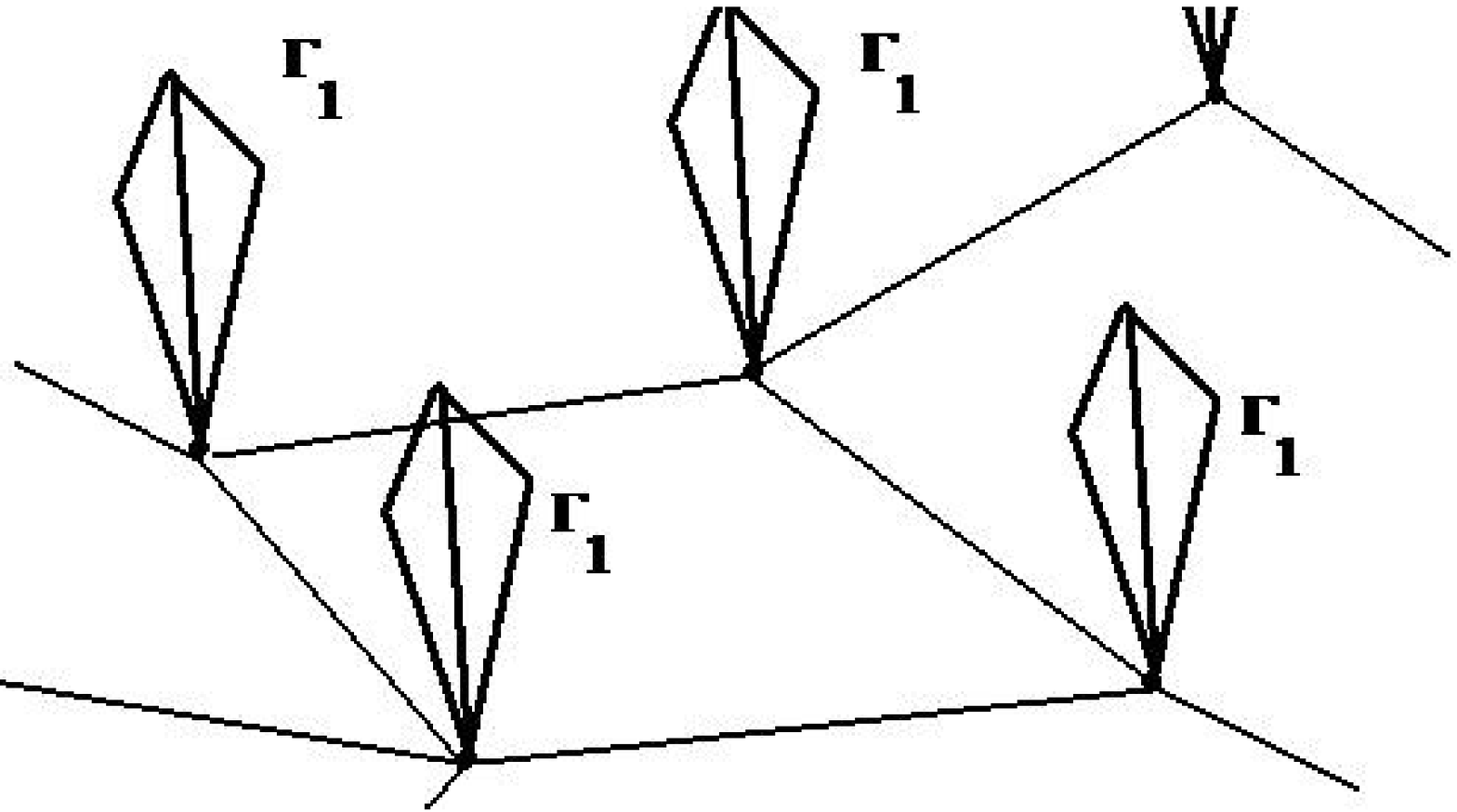}\includegraphics{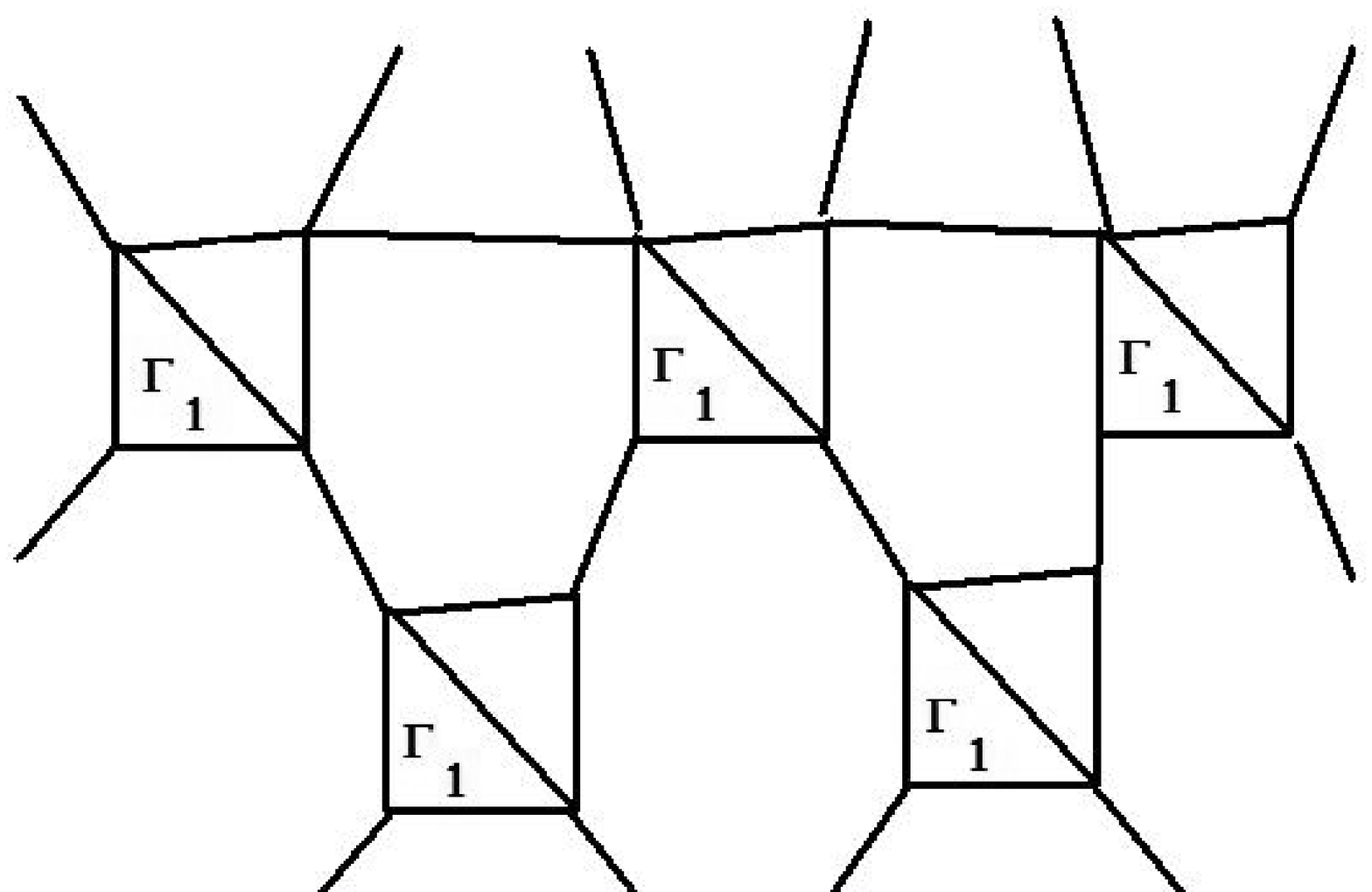}} \end{center}
\caption{Two types of ``decorations'' of a graph $\Gamma_0$. On the left, a copy of $\Gamma_1$ is
attached to each vertex of $\Gamma_0$. In the one shown on the right,
each vertex of a {\bf regular} graph $\Gamma_0$ is replaced by a fixed graph $\Gamma_1$.}
\label{F:decor}
\end{figure}

Another statement that does not depend on the specific structure of an infinite graph is the
``decoration'' method of gap opening, due to \cite{AiS} (in the combinatorial case)\footnote{Some
indications of the presence of this effect were discussed previously in \cite{AEL,Pav2}}. Assume
that one ``decorates'' a combinatorial or quantum, finite or infinite graph $\Gamma_0$
by attaching to its each vertex a fixed finite graph $\Gamma_1$ (Fig. \ref{F:decor}).
Then, under mild conditions, the spectrum of the so decorated graph has mandatory gaps near the
eigenvalues of the decoration. The reason is that one can eliminate the decorations, as it was
described before, by replacing them with their Dirichlet-to-Neumann maps. These maps depend on the
spectral parameter in a singular way: they are meromorphic with poles on the spectrum of the
decoration. This introduces a vertex ``potential'' that blows up at some points, thus preventing
the spectrum from appearing near these points. One can find the details for the combinatorial case
in \cite{AiS} and for the quantum one in \cite{Ku_graphs2}.

These are probably the only general results about spectra of infinite quantum graphs that are known to the author. However, for special subclasses of infinite graphs, a more detailed analysis is possible, which usually dwells on some kind of a symmetry.

\subsubsection{Radial trees} In the case when the graph is a rooted tree, whose properties (degrees of vertices and lengths of edges) depend only on the distance from the root, a simple harmonic analysis is possible, which essentially reduces the problem to a weighted ODE problem on the half line. One can find the corresponding results in  \cite{SoSo,Solom,Solom03,NS,NS2}.

\subsubsection{Periodic graphs} If an abelian group acts freely and co-compactly on the graph,
abelian harmonic analysis (akin to the standard Floquet theory for PDEs \cite{Ku93,RS,Eastham})
is possible, which proves completeness of the so called Floquet-Bloch generalized eigenfunctions,
the band-gap structure of the spectrum, Liouville type theorems, etc. There are some differences
with the continuous case, though. For instance, while in the periodic elliptic PDE case the spectrum
is absolutely continuous \cite{Ku93,RS}, this is not true anymore in the graph case, where compactly
supported eigenfunctions arise \cite{Ku89,Ku_graphs2}. Also, localized perturbations of periodic
structures and operators can lead to existence of eigenvalues embedded into the continuous spectrum,
which does not happen in the PDE case.
One can find these and other discussions of  periodic graph problems in
\cite{ExTaVa,Ex4,Ex6,Below_periodic,Fox,AEL,Ku89,KuchPin,Ku01,Ku_graphs2,HKSW,Pavl2,KK99,KK02,KuchPost}.

\subsubsection{Other classes of infinite graphs} There are studies of spectral properties on some other classes of infinite combinatorial and quantum graph structures, such as the so called limit operators, operators of quasi-crystal type, random operators, etc. (e.g., \cite{Rab,Linz,Exner_loc,AizSimsWarz1,AizSimsWarz2,AizSimsWarz3,AizSimsWarz4,AizSimsWarz5,HisPost} and references therein).

\subsection{Inverse problems} One can ask the natural question, analogous to the famous problem of spectral geometry ``Can one hear the shape of the drum?'' In the quantum graph setting, the question is of whether one can reconstruct the topology, vertex conditions, and potentials of the Hamiltonian of a finite quantum graph from the spectral data, or, if some infinite leads are attached, from scattering data. It is known that in general this is impossible (see  \cite{KuS,Below,KurBom}). However, the beautiful construction of \cite{GutSmi} shows that ``one can hear the shape of a quantum graph'' if one assumes rational independence of edges' lengths.
See also \cite{Bel2,Bel3,C3,KurNow,Freiling,Piv,Yurko} for the control theory approach and other discussions of the inverse problem.

\subsection{Nodal domains} An important part of spectral theory of differential operators is studying
nodal domains of eigenfunctions of a differential operator with discrete spectrum. In dimension one,
i.e. on a finite interval, this is done by the well known oscillation theorems that essentially claim
that the $n$th eigenfunction has $n$ nodal domains. In higher dimensions, this becomes an upper bound
(and not a sharp one) for the number of nodal domains \cite{CH,Henrot}. Many ``simplest'' questions
still do not have their answers, e.g. that the number of nodal domains cannot be bounded over the
whole spectrum, or that the nodal set for the second eigenfunction in a simply-connected domain always
hits the boundary (Payne conjecture). Many researchers have been interested in studying nodal domains
of eigenfunctions on combinatorial or quantum graphs. Recently, counting of nodal domains was suggested
as a tool to resolve the problem of isospectrality of non-isomorphic graphs. See
\cite{LNM_nodal,Berk_nodal} and \cite{Band} in \cite{volume} for surveys, recent results, and references.

\subsection{Relations to discrete operators}

This introductory survey is devoted to the new area of quantum graphs and their spectra. On the other
hand, the combinatorial counterpart of this theory, which is sometimes called {\bf discrete geometric
analysis}, is a rather well established topic (see e.g., books \cite{Sunada_book,Biggs,Chung,CDS,CDS2,Col} and
the surveys \cite{Sunada,MW,Terras}). One wonders whether there is a relation between the two. The
answer is a ``yes,'' although the relation is not always straightforward.

We will show now how spectral problems for quantum graphs can
sometimes be transformed into the ones for difference operators on
combinatorial graphs. This observation goes back at least to \cite{Al,deG}
(see also \cite{Ex7,Pankr,Cat,Ku_graphs2} for more detailed
considerations).

Let us consider the simplest case of $\Gamma$ a finite graph (the procedure works for infinite graphs as well, but needs to be done more carefully \cite{Pankr}) with the Hamiltonian ${\mathcal H}$ defined as the negative second derivative with Neumann-Kirchhoff conditions
(\ref{E:delta0_cond}). Since the spectrum $\sigma({\mathcal H})$ is
discrete, we need to look for eigenfunctions, i.e. solutions of the equation
\begin{equation}\label{E:sp_pr_fin}
{\mathcal H}f=\lambda f
\end{equation}
with $f \in L_2(\Gamma)$. Let $v$ be a vertex and $e$ one of
the outgoing edges of length $l_e$ with the coordinate $x$
counted from $v$. Let us denote by $w_e$ the other end of
$e$. Then along this edge one can solve (\ref{E:sp_pr_fin}):
\begin{equation}\label{E:alongedge}
f_e(x)=\frac{1}{\sin{\sqrt{\lambda}l_e}}\left(f_e(v)\sin{\sqrt{\lambda}(l_e-x)}
+f_e(w_e)\sin{\sqrt{\lambda}x} \right).
\end{equation}
This can be done as long as $\lambda \neq n^2\pi^2l_e^{-2}$ with
an integer $n\neq 0$ (the formula can be interpreted for
$\lambda=0$), i.e. when $\lambda$ does not belong to the spectrum
of the operator on the edge $e$ with Dirichlet conditions at the edge's ends.

The last formula allows us to find the derivative at $v$:
\begin{equation}\label{E:deriv_alongedge}
f_e^\prime
(v)=\frac{\sqrt{\lambda}}{\sin{l_e\sqrt{\lambda}}}\left(f_e(w_e)-f_e(v)\cos{l_e\sqrt{\lambda}}
\right).
\end{equation}
Substituting these relations into (\ref{E:delta0_cond}) to
eliminate the derivatives, one reduces (\ref{E:delta0_cond})
to a system of discrete equations that involve only the vertex
values:
\begin{equation}\label{E:discr_finite}
T(\lambda)F=0.
\end{equation}
Here $T(\lambda)$ is a square matrix of dimension $\sum\limits_v d_v$.

One can notice that (\ref{E:discr_finite}) is a system of
second order \textbf{difference} equations on the {\bf combinatorial}
version of the graph $\Gamma$, where at each vertex $v$ we have a
$d_v$-dimensional value $F(v)$ of the vector function $F$
assigned. The components of this vector are labeled by the edges
adjacent to $v$. Notice that if the graph is not regular (i.e., if
$d_v$ is not constant throughout $\Gamma$), then the dimensions of
the vector values are changing from vertex to vertex.
One concludes that the following statement holds:
\begin{theorem}\label{T:con_disc_fin}
A point $\lambda \neq n^2\pi^2l_e^{(-2)}$ belongs to the spectrum
of ${\mathcal H}$ if and only if zero belongs to the spectrum of the
matrix pencil $T(\lambda)$.
\end{theorem}

This theorem shows that spectral problems for quantum graph
Hamiltonians can be rewritten as spectral problems for some
difference operators on the combinatorial counterpart of the graph.
In general, though, (\ref{E:discr_finite}) might look complicated. It
simplifies significantly for some frequently arising situations.
Indeed, if we assume that all edges are of same
length $l$, then (\ref{E:discr_finite}) becomes at each vertex
\begin{equation}
\label{E:discr_version}
\sum\limits_{e=(v,w)
\in E_v} {f_e(w_e)}={\cos{l\sqrt{\lambda}}}F(v).
\end{equation}
This means that $\lambda \neq n^2\pi^2l^{(-2)}$ belongs to the spectrum of the quantum graph Hamiltonian if and only if $\cos{l\sqrt{\lambda}}$ belongs to the spectrum of the discrete Laplace operator that maps $\{f(v)\}$ into $\{\sum\limits_{e=(v,w)
\in E_v} {f_e(w_e)}\}$. This enables one to transfer known spectral results from discrete geometric analysis to the quantum graph situation. In the case of infinite graphs, though, the question arises of whether the various types (pure point, singular continuous, absolutely continuous) of the spectrum are preserved under this transformation. The positive answer can be found in \cite{Pankr}.

The author would like to take this opportunity to mention the often forgotten important paper
\cite{Shubin_pseudodiff}, where a discrete analog of pseudodifferential operator technique is
developed with important applications to spectral theory and Greens function estimates in discrete
setting.

\section{PDE and Control problems}

Problem of boundary control of partial differential equations on graphs has been considered by
several authors, due to many applications in engineering. Here as the
boundary of a graph one considers its vertices of degree $1$. The bottom line of these studies
can be roughly summarized as follows: presence of cycles in a graph prohibits controllability,
so one tries to apply control on trees only; on trees, one should control either all, or all but
one boundary vertices. One can find the details and references in \cite{Zua,LLS,Nic3,Nic4,Bel1}
and in the survey \cite{Avdonin} in \cite{volume}.

A variety of other PDE problems on quantum graphs have been considered, see details and references
in \cite{AM1,AM2,AM3,Be2,Be3,Penkin,Pokornyi,Mehmeti,KoSsemigr,C1,C2,Carlson4,Nic2}.

\section{Various generalizations of quantum graphs}

Many generalizations of quantum graphs have been studied due to various applications. The operators
we have considered so far were scalar (i.e., spin was not taken into account). However,
\textbf{matrix operators} such as Dirac operators and Rashba Hamiltonians have also been considered,
see the survey \cite{Harrison} in \cite{volume} for details and references.

\textbf{Differential operators of orders higher than $2$} were considered, due to the needs of
photonic crystal theory, in \cite{KK02}.

Both photonic crystal theory and quantum waveguide theory have lead (albeit in somewhat different
manner) to the necessity of considering the problems, where the particle is not strictly confined
to a graph, but rather attracted to it by a delta-type potential along the graph. This, in
particular,
allows for tunneling between distant parts of the graph, rather than forcing the quantum particle
to move through the vertices, as in the standard quantum graph theory. This explains why these
systems are sometimes called {\bf leaky graphs}. One can find a detailed survey on leaky graphs in
\cite{Exner} in \cite{volume} and considerations of such systems as coming from the photonic crystals
in \cite{FK2,FK4,KK99,KK02}. The leaky graph operators can usually be written as
``pseudo-differential'' operators of first order on graphs \cite{KK02,Paiv,FK4}.

In many applications, it is interesting to study analogs of quantum graphs that consist of cells
of
a higher dimension, for instance branching surface structures (sometimes called ``open book
structures'').
Such necessity arises in dynamical systems, fluid dynamics, as well as in photonic crystal theory.
It is also interesting to allow combinations of cells of different dimension, e.g. one-dimensional
edges attached to two-dimensional surfaces or three-dimensional volumes. These are the so called
\textbf{multistructures}. One can find some information on such objects in
\cite{Freidlin_openbook,Pokornyi,Mehmeti,MNP,ExS_mult,Brun_horns,Brun_exner,ExSantenna}, although
the theory here is not nearly as well developed as for quantum graphs.

Another interesting direction related to quantum graphs both in spirit and in terms of some shared
techniques, is \textbf{analysis on fractals}. One can find basic introduction to this analysis and
references in \cite{Kigami,Strichartz} and a survey of some exciting recent developments in
\cite{Nekr} in \cite{volume}.

\section{Acknowledgment\label{S: A}}

The author planned to express his gratitude to all people who over the years helped him to study
and progress through the quantum graph theory. He discovered, however, that the list would be so
long, that this would increase significantly the length of the text. It has thus become necessary
to express a blanket gratitude to numerous friends and colleagues, discussions with and information
from whom were instrumental in writing this article. Particular thanks go to the Editors Pavel
Exner, Jon Keating, Toshikazu Sunada, and Alex Teplyaev, who have spent a lot of time working on
this volume. Part of this work was done when the author was at
the Isaac Newton Institute for Mathematical Sciences on sabbatical leave from the Texas A\&M
University. The author expresses his gratitude to INI and TAMU for this support.

This research was partly sponsored by the NSF through the DMS
Grants 9971674, 0072248, 0296150, and 0648786. The author is thankful to the NSF for the support.

The author realizes that the list of references must be severely incomplete and apologizes for this
incompleteness. He hopes that the pointers to many survey papers provided here will enable the
reader to find further references.



\begin{thebibliography}{999}

\bibitem{AizSimsWarz1} M.~Aizenman, R.~Sims, S.~Warzel, Fluctuation based proof
of the stability of ac spectra of random operators on tree graphs,
in \cite{BCFK}, 1--14.

\bibitem{AizSimsWarz2} M.~Aizenman, R.~Sims, S.~Warzel, Fluctuation Based Proof
of the Stability of AC Spectra of Random Operators on Tree Graphs.
In Recent Advances in Differential Equations and Mathematical
Physics, N. Chernov, Y. Karpeshina, I.W. Knowles, R.T. Lewis, and
R. Weikard (eds.). AMS Contemporary Mathematics Series (to appear
in 2006). (http://arxiv.org/math-ph/0510069)

\bibitem{AizSimsWarz3} M.~Aizenman, R.~Sims, S.~Warzel, Persistence Under Weak Disorder of
AC Spectra of Quasi-Periodic Schr\"{o}dinger Operators on Trees
Graphs. To appear in: Moscow Mathematical Journal.
(http://arxiv.org//math-ph/0504084)

\bibitem{AizSimsWarz4} M.~Aizenman, R.~Sims, S.~Warzel, Absolutely Continuous Spectra of
Quantum Tree Graphs with Weak Disorder, Comm. Math. Phys. 264,
371 389 (2006)

\bibitem{AizSimsWarz5} M.~Aizenman, R.~Sims, S.~Warzel, Stability of the absolutely
continuous spectrum of random Schr\"{o}dinger operators on tree
graphs. Probab. Theory Relat. Fields , 136 no.3, 363394 (2006)

\bibitem{AkhGlaz} N. Akhiezer and I. Glazman, \textit{Theory of Linear
Operators in Hilbert Space}, Dover, NY 1993.

\bibitem{ACD}  E. Akkermans, A. Comtet, J. Desbois, G. Montambaux, C.
Texier, Spectral determinant on quantum graphs, Ann. Physics 284
(2000), no. 1, 10--51.

\bibitem{Al}  S. Alexander, Superconductivity of networks. A percolation
approach to the effects of disorder, Phys. Rev. B, 27 (1983),
1541-1557.

\bibitem{amovilli:04}C.~Amovilli, F.~Leys and N.~March, Electronic
  energy spectrum of two-dimensional solids and a chain of C atoms
  from a quantum network model, Journal of Math. Chemistry, {\bf 36} (2004),
  no. 2, 93--112.

\bibitem{Avdonin} S.~Avdonin, Control Problems on Quantum Graphs, in \cite{volume}.

\bibitem{Avr}  J. Avron, Adiabatic quantum transport, in E. Akkermans,
G. Montambaux, J.-L. Pichard (Editors), \textit{%
Physique Quantique M\'{e}soscopique. Mesoscopic Quantum Physics
(Les Houches Summer School Proceedings)}, Vol 61, Elsevier Science
Publ. 1995, 741-791.

\bibitem{AEL}  J. Avron, P. Exner, and Y. Last, Periodic Schr\"{o}dinger
operators with large gaps and Wannier-Stark ladders, Phys. Rev.
Lett. 72(1994), 869-899.

\bibitem{ARZ}  J. Avron, A. Raveh, and B. Zur, Adiabatic quantum
transport in multiply connected systems, Rev. Mod. Phys. 60(1988),
no.4, 873-915.

\bibitem{AKK}  W. Axmann, P. Kuchment, and L. Kunyansky, Asymptotic
methods for thin high contrast 2D PBG materials, J. Lightwave Techn.,
17(1999), no.11, 1996- 2007.

\bibitem{Band} R.~Band, I.~Oren, and U.~Smilansky, Nodal domains on graphs
- How to count them and why?, in \cite{volume}.


\bibitem{Bel1} M.~I.~Belishev, On the boundary controllability of a dynamical system
described by the wave equation on a class of graphs (on trees),
J. Math. Sci. (N. Y.), 132, no. 1, 11-25, 2004.

\bibitem{Bel2} M.~I.~Belishev, Boundary spectral inverse problem on a class of
graphs (trees) by the BC method, Inverse Problems, {\bf 20}
(2004), 647--672.

\bibitem{Bel3} M.~I.~Belishev and A.~F.~Vakulenko, Inverse problems
on graphs: recovering the tree of strings by the BC-method, {\it
J. Inv. Ill-Posed Problems}, {\bf 14} (2006), 29--46.


\bibitem{Mintchev} B. Bellazzini, M. Burrello, M. Mintchev, and P. Sorba,
Quantum Field Theory on Star Graphs, in \cite{volume}.


\bibitem{Be2}  J. von Below,
Sturm-Liouville eigenvalue problems on networks,  Math. Meth.  Appl.
Sc., {\bf 10}, 1988, 383-395.

\bibitem{Be3} J. von Below, {\it Parabolic Network Equations}, Habilitation
Thesis, Eberhard-Karls-Universit\"at T\"ubingen, 1993.

\bibitem{Below_periodic} J.~von Below, The index of a periodic graph, Results Math. {\bf 25} (1994), no. 3-4, 198--223.

\bibitem{Below} J.~von Below, Can one hear the shape of a network?, in \cite{Mehmeti} 19--36, 2001.

\bibitem{Berez} Yu.~Berezanskii, \textit{Expansions in Eigenfunctions of Selfadjoint
Operators}, American Mathematical Society, Providence, 1968.

\bibitem{Berk} G.~Berkolaiko, Two constructions of quantum graphs and two types of spectral statistics, in \cite{volume}.

\bibitem{Berk_nodal} G. Berkolaiko, A lower bound for nodal count on discrete and metric
graphs, to appear in Commun. Math. Phys.

\bibitem{BCFK} G.~Berkolaiko, R.~Carlson, S.~Fulling, and P.~Kuchment
(Editors), Quantum Graphs and Their Applications, Contemp. Math.,
v. 415, American Math. Society, Providence, RI 2006.

\bibitem{BHW} G.~Berkolaiko, J.~Harrison \& J.~Wilson,
Mathematical aspects of vacuum energy on quantum graphs,
Preprint arXiv:0711.2707.

\bibitem{Biggs} N.~Biggs, \textit{Algebraic Graph Theory},
Cambridge Univ. Press 2001.


\bibitem{LNM_nodal} T. Biyikoglu, J. Leydold, and P.F. Stadler,
\textit{Laplacian Eigenvectors of Graphs: Perron-Frobenius and Faber-Krahn Type Theorems},
Lecture Notes in Mathematics , Vol. 1915, 2007.

\bibitem{Exner_book} J. Blank, P. Exner, and M. Havl\'{i}\v{c}ek,
\textit{HilbertSpace Operators in Quantum Physics}, 2nd edition, Springer Verlag 2008.

\bibitem{Bolte} J.~Bolte and S.~Endres, Trace formulae for quantum graphs, in \cite{volume}.

\bibitem{KurBom} J.~Boman and P.~Kurasov, Symmetries of quantum graphs and the inverse
scattering problem,
Adv. in Appl. Math. {\bf 35} (2005), no. 1, 58--70.

\bibitem{BridsVogt} M.~R.~Bridson and K.~Vogtmann, The symmetries of outer space, Duke Math. J.
{\bf 106}, no. 2, 391--409.

\bibitem{Brun_exner}J. Br\"{u}ning, P. Exner, and V.A. Geyler, Large gaps in point-coupled periodic
systems of manifolds, J. Phys. A: Math. Gen. {\bf 36} (2003), 4875--4890.

\bibitem{Brun_horns}J. Br\"{u}ning and V.~A.~Geyler, Scattering on compact manifolds with
in¯nitely thin horns, J. Math. Phys. {\bf 44} (2003), 371--405.

\bibitem{Brun_qsphere}J. Br\"{u}ning, V.~A.~Geyler, V.A. Margulis and M.A. Pyataev, Ballistic con-
ductance of a quantum sphere, J. Phys. A {\bf 35} (2002), 4239--4247.

\bibitem{Brun_cantor}J. Br\"{u}ning, V.~A.~Geyler, and K.~Pankrashkin,
Cantor and Band Spectra for Periodic Quantum Graphs with Magnetic Fields,
Comm. Math. Phys. {\bf 269} (2007), no. 1, 87--105.

\bibitem{Brun_extension}J. Br\"{u}ning, V.~A.~Geyler, and K.~Pankrashkin,
Spectra of self-adjoint extensions and applications to solvable Schr\"{o}dinger operators,
Rev. Math. Phys. {\bf 20} (2008), 1--70.


\bibitem{BT}  W. Bulla and T. Trenkler, The free Dirac operator on
compact and non-compact graphs, J. Math. Phys. 31(1990), 1157 - 1163.

\bibitem{Buttiker} M. B\"{u}ttiker,
Absence of backscattering in the quantum Hall effect in multiprobe
conductors, Phys. Rev. B {\bf 38} (1988), 9375--9389.

\bibitem{acoustics} C. Cacciapuoti, R. Figari, and A. Posilicano, Point interactions in acoustics:
One-dimensional models, J. Math. Phys. 47 (2006), 062901.

\bibitem{C1}  R. Carlson, Hill's equation for a homogeneous tree,
Electronic J. Diff. Equations 1997 (1997), no.23, 1-30.

\bibitem{C2}  R. Carlson, Adjoint and self-adjoint operators on graphs,
Electronic J. Diff. Equations (1998), no.6, 1-10.

\bibitem{C3}  R. Carlson, Inverse eigenvalue problems on directed
graphs, Trans. Amer. Math. Soc. 351 (1999), no. 10, 4069--4088.

\bibitem{Carlson4}  R.~Carlson, Nonclassical Sturm-Liouville problems and
Schr\"{o}dinger operators on radial trees, Electronic J. Diff.
Equat, 2000 (2000), no.71, 1-24.

\bibitem{Carlson5} R.~Carlson, Spectral theory and spectral gaps for periodic
Schrodinger operators on product graphs, in \cite{WRM}, S29--S46, 2004.

\bibitem{Carlson} R.~Carlson, Boundary value problems on infinite metric graphs,
in \cite{volume}.

\bibitem{Carlson_lung} R.~Carlson, Linear Network Models Related to Blood Flow, in \cite{BCFK},
65--80, 2006.


\bibitem{Cat}  C. Cattaneo, The spectrum of the continuous Laplacian on
a graph, Monatsh. Math. 124 (1997), no. 3, 215--235.

\bibitem{Cheon}T.~Cheon and P.~Exner, An approximation to $\delta'$ couplings on graphs,
J. Phys. A {\bf 37} (2004), no. 29, L329--L335.

\bibitem{Chung}  F. Chung, \textit{Spectral Graph Theory}, Amer. Math.
Soc., Providence R.I., 1997.

\bibitem{Colin}  Y. Colin de Verdi\`{e}re, Sur la multiplicit¶e de la premi\`{e}re valeur
propre non nulle du Laplacien, Comment. Math. Helv., {\bf 61} (1986), 254--270.

\bibitem{Col}  Y. Colin de Verdi\`{e}re, \textit{Spectres De Graphes},
Societe Mathematique De France, 1998



\bibitem{CH} R.~Courant and D.~Hilbert, \textit{Methods of
Mathematical Physics, Volume II Partial Differential Equations},
Interscience, New York, 1962.

\bibitem{CullVogt} M.~Culler and K.~Vogtmann, Moduli of graphs and
automorphisms of free groups, Inventiones Math. {\bf 84} (1986), 91--119.

\bibitem{CDS}  D. Cvetkovic, M. Doob, and H. Sachs, \textit{Spectra of
Graphs}, Acad. Press., NY 1979.

\bibitem{CDS2}  D. Cvetkovic, M. Doob, I. Gutman, A. Targasev, \textit{%
Recent Results in the Theory of Graph Spectra}, Ann. Disc. Math.
36, North Holland, 1988.

\bibitem{Cycon} H.~L.~Cycon, R.~G.~Froese, W.~Kirsch, and B.~Simon,
\textit{Schr\"{o}dinger Operators: With Application to Quantum
Mechanics and Global Geometry}, Texts and Monographs in Physics,
Springer Verlag 1987.

\bibitem{Zua} R. Dager, E. Zuazua {\it Wave propagation, observation and control in
$1\text{-}d$ flexible multi-structures.}
Mathematiques and Applications (Berlin), 50, Springer-Verlag,
Berlin, 2006.

\bibitem{deG}  P.-G. de Gennes, Champ critique d'une boucle supraconductrice
ramefiee, C. R. Acad. Sc. Paris 292B (1981), 279-282.


\bibitem{Nic3} B. Dekoninck and S. Nicaise,   Control of
 networks of Euler-Bernoulli beams,
ESAIM-COCV, {\bf 4} (1999), 57--82.


\bibitem{Eastham}  M.~S.~P.~Eastham, \textit{The Spectral Theory of Periodic
Differential Equations}, Scottish Acad. Press Ltd.,
Edinburgh-London, 1973.

\bibitem{EH}  W. D. Evans and D. J. Harris, Fractals, trees and the
Neumann Laplacian, Math. Ann. 296(1993), 493-527.


\bibitem{evans2}
W.D.\,Evans and Y.\,Sait\=o,  Neumann Laplacians on domains and
operators on associated trees, Q. J. Math.,  {\bf 51}
(2000), 313--42.

\bibitem{Everitt1} W.~Everitt and L.~Markus, \textit{Boundary Value
Problems and Symplectic Algebra for Ordinary Differential and
Quasi-Differential Operators}, AMS 1999.

\bibitem{Everitt2} W.~Everitt and L.~Markus, Complex symplectic geometry
with applications to ordinary differential equations, Trans. AMS
{\bf 351} (1999), 4905--4945.

\bibitem{Everitt3} W.~Everitt and L.~Markus, \textit{Multi-Interval Linear
Ordinary Boundary Value Problems and Complex Symplectic Algebra},
Memoirs of the AMS, No. 715, 2001.

\bibitem{Ex4}  P. Exner, Lattice Kronig-Penney models, Phys. Rev. Lett.
74 (1995), 3503-3506.

\bibitem{Ex5}  P. Exner, Weakly coupled states on branching graphs,
Lett. Math. Phys. {\bf 38} (1996), 313-320

\bibitem{Ex6}  P. Exner, Contact interactions on graph superlattices, J.
Phys. A29 (1996), 87-102

\bibitem{Ex7}  P. Exner, A duality between Schroedinger operators on
graphs and certain Jacobi matrices, Ann. Inst. H. Poincare 66
(1997), 359-371.

\bibitem{Ex8}  P. Exner, Bound states of infinite curved polymer chains,
Lett. Math. Phys. 57 (2001), 87-96

\bibitem{Exner} P. Exner, Leaky quantum graphs: a review, in \cite{volume}.

\bibitem{EG}  P. Exner, R. Gawlista, Band spectra of rectangular graph
superlattices, Phys. Rev. B53 (1996), 7275-7286.

\bibitem{Exner_loc} P.~Exner, M.~Helm, and P.~Stollmann, Localization on a
quantum graph with a random potential on the edges, Rev. Math. Phys. {\bf 19} (2007),
no. 9, 923--939.

\bibitem{volume} P.~Exner, J.~P.~Keating, P.~Kuchment, T.~Sunada, and A.~Teplyaev (Editors),
\textit{Analysis on Graphs and its Applications}, Proc. Symp. Pure Math., AMS, to appear.

\bibitem{ExKov}P.~Exner and H.~Kova\v{r}\'{i}k, Magnetic strip waveguides, J. Phys.
A {\bf 33} (2000), no. 16, 3297--3311.

\bibitem{ExPost} P.~Exner and O.~Post, Convergence of spectra of graph-like thin manifolds,
J. Geom. Phys. {\bf 54} (2005), no. 1, 77--115.

\bibitem{ExPost2} P.~Exner and O.~Post, Convergence of resonances on thin branched
quantum waveguides. J. Math. Phys. 48 (2007), no. 9, 092104, 43 pp.

\bibitem{ExS_mult}  P. Exner and P. \v {S}eba, Quantum
motion on a half-line connected to a plane, J. Math. Phys. {\bf 28} (1987),
386--391.

\bibitem{ExS1}  P. Exner and P. \v {S}eba, Electrons in semiconductor
microstructures: a challenge to operator theorists, in \textit{%
Schr\"{o}dinger Operators, Standard and Nonstandard (Dubna
1988)}., World Scientific, Singapore 1989; pp. 79-100.


\bibitem{ExS3}  P. Exner, P. \v {S}eba, Free quantum motion on a
branching graph, Rep. Math. Phys. 28 (1989), 7-26.

\bibitem{ExS2}  P. Exner and P. \v {S}eba,  Bound states in curved
quantum waveguides, J. Math. Phys. {\bf 30} (1989), 2574--2580.

\bibitem{ExSantenna}  P. Exner and P. \v {S}eba, Resonance statistics in a microwave cavity with
a thin antenna, Phys. Lett. A {\bf 228} (1997), 146--150.

\bibitem{ExTaVa} P. Exner, M. Tater, D. Van\v{e}k, A single-mode quantum transport
in serial-structure geometric scatterers, J. Math. Phys. {\bf 42} (2001),
4050--4078.

\bibitem{Exner_cond} P. Exner and O.~Turek, Approximations of singular vertex
couplings in quantum graphs, Rev. Math. Phys. {\bf 19} (2007), no. 6, 571--606.

\bibitem{ExVug} P.~Exner and S.~A.~Vugalter, On the number of particles that a
curved quantum waveguide can bind, J. Math. Phys. {\bf 40} (1999), no. 10, 4630--4638.


\bibitem{FK2}  A. Figotin and P. Kuchment, Band-gap structure of the
spectrum of periodic and acoustic media. II. 2D Photonic crystals,
SIAM J. Applied Math. 56(1996), 1561-1620.

\bibitem{FK4}  A. Figotin and P. Kuchment, Spectral properties of
classical waves in high contrast periodic media, SIAM J. Appl. Math.
58(1998), no.2, 683-702.

\bibitem{Gnu_Andreev}H.~Flechsig and S.~Gnutzmann, On the spectral gap in Andreev
graphs, in \cite{volume}.

\bibitem{Fox} C.~Fox, V.~Oleinik, and B.~Pavlov, A Dirichlet-to-Neumann map approach
to resonance gaps and bands of periodic networks, in \textit{ Recent advances in
differential equations and mathematical physics}, 151--169, Contemp. Math., 412, Amer.
Math. Soc., Providence, RI, 2006.

\bibitem{Freidlin}  M. Freidlin, \textsl{Markov Processes and Differential
Equations: Asymptotic Problems}, Lectures in Mathematics ETH
Z\"{u}rich, Birkh\"{a}user Verlag, Basel, 1996.

\bibitem{Freidlin2} M.~Freidlin, M.~Weber, Small Diffusion Asymptotics for Exit
Problems on Graphs, in \cite{BCFK}, 137--150.

\bibitem{Freidlin3}  M. Freidlin and A. Wentzell, Diffusion processes on graphs and
the averaging principle, Annals of Probability, 21(1993), no.4,
2215-2245.

\bibitem{Freidlin_openbook} M. Freidlin and A. Wentzell, Diffusion processes on an open
book and the averaging principle, Stochastic Process. Appl. {\bf 113} (2004), no. 1, 101--126.

\bibitem{Freiling} G. Freiling, M.Yu. Ignat'ev, and V.A. Yurko, An inverse spectral
problem for Sturm-Liouville operators with singular potentials on star-type graphs,
in \cite{volume}.

\bibitem{Fr_graph} L.~Friedlander, Genericity of simple eigenvalues
for a metric graph, Israel J. Math. 146 (2005), 149--156.

\bibitem{Fr_graph2} L.~Friedlander, Extremal properties of eigenvalues for a
metric graph, Ann. Inst. Fourier (Grenoble) 55 (2005), no. 1,
199--211.

\bibitem{Fr_graph_det} L.~Friedlander, Determinant of the
Schr\"{o}dinger operator on a metric graph, in \cite{BCFK},
151--160.


\bibitem{Fulling_vac} S.~A.~Fulling, Local spectral density and vacuum energy near
a quantum graph vertex, in \cite{BCFK}, 161--172, 2006.

\bibitem{Fulling_vac2} S.~A.~Fulling, Vacuum energy and spectral analysis for Robin
boundaries and quantum graphs, J. Phys. A {\bf 39} (2006), no. 21, 6377--6383.

 \bibitem{Fulling2} S. A. Fulling, L. Kaplan, and J. H. Wilson, Vacuum Energy and Repulsive
 Casimir Forces in Quantum Star Graphs, Phys. Rev. A {\bf 76} (2007), 012118.

\bibitem{FKW}S.~Fulling, P.~Kuchment, and J. Wilson, Index theorems for quantum graphs,
J. Phys. A: Math. Theor. {\bf 40} (2007), 14165--14180.

\bibitem{Fulling} S.~Fulling and J. Wilson, Vacuum Energy and Closed Orbits in Quantum
Graphs, in \cite{volume}.

\bibitem{G}  N. Gerasimenko, Inverse scattering problem on a noncompact
graph, Theor. and Math. Phys. 75 (1988), 460--470.

\bibitem{GP}  N. Gerasimenko and B. Pavlov, Scattering problems on
non-compact graphs, Theor. Math. Phys., 74 (1988), no.3, 230--240.



\bibitem{Glaz} I.~M.~Glazman, \textit{Direct Methods of Qualitative
Spectral Analysis of Singular Differential Operators}, Isr. Progr.
Sci. Transl., Jerusalem 1965.

\bibitem{GnuSmi}S. Gnutzmann and U. Smilansky, Quantum graphs: Applications
to quantum chaos and universal spectral statistics, Advances in Physics,
{\bf 55} (2006), no. 5, 527–-625.

\bibitem{Gri_survey} D.~Grieser, Thin tubes in mathematical physics, global
analysis and spectral geometry, in \cite{volume}.

\bibitem{Gri1}  J.S. Griffith, A free-electron theory of conjugated
molecules. I. Polycyclic Hydrocarbons, Trans. Faraday Soc., 49
(1953), 345-351.

\bibitem{Gri2}  J.S. Griffith, A free-electron theory of conjugated
molecules. II. A derived algebraic scheme, Proc. Camb. Philos.
Soc., 49 (1953), 650-658.

\bibitem{GrigNekr} R.~Grigorchuk and V.~Nekrashevych, Self-similar groups, operator
algebras and Schur complement, J. Modern Dynamics {\bf 1} (2007), no. 3, 323--370.

\bibitem{Linz}M.~J.~Gruber, D.~H.~Lenz, and I.~Veseli\'{c}, Uniform existence of the
integrated density of states for models om $\ZZ^2$, in \cite{volume}.

\bibitem{GutSmi} B.~Gutkin and U.~Smilansky, Can one hear the shape of a graph?,
J. Phys. A. Math. Gen. {\bf 34} (2001), 6061--6068.

\bibitem{Har}  M. Harmer, Hermitian symplectic geometry and extension
theory, J. Phys. A: Math. Gen. 33(2000), 9193-9203.

\bibitem{Harrison} J.~M.~Harrison, Quantum graphs with spin Hamiltonians,
in \cite{volume}.

\bibitem{HKSW}J. M. Harrison, P. Kuchment, A. Sobolev, and B. Winn. On occurrence of spectral
edges for periodic operators inside the Brillouin zone. Journal of Physics
A: Mathematical and Theoretical, {\bf 40(27)} (2007),7597-–7618.

\bibitem{Henrot}A.~Henrot, \textit{Extremum Problems for Eigenvalues of Elliptic Operators},
Frontiers in Mathematics, Birkh\"{a}user, 2006.

\bibitem{HisPost} P.~Hislop and O.~Post, Anderson Localization for radial tree-like
random quantum graphs, arXiv:math-ph/0611022.


\bibitem{Terras} M.~D.~Horton, H.~M.~Stark, and A.~A.~Terras, Zeta functions of
weighted and covering graphs, in \cite{volume}.

\bibitem{Sirko} O.~Hul, M.~{\L}awniczak, S.~Bauch, and L.~Sirko, Simulation of
quantum graphs by microwave networks, in \cite{volume}

\bibitem{Hurt} N.E. Hurt, \textit{t Mathematical Physics of Quantum Wires and
Devices}, Kluwer Acad. Publ., 2000.

\bibitem{Keating} J.~P.~Keating, Quantum graphs and quantum chaos, in \cite{volume}.

\bibitem{Kigami} J. Kigami, \textit{Analysis on Fractals},
Cambridge Univ. Press 2001.

\bibitem{korotyaev:06_magn} E~Korotyaev and I.~Lobanov, Zigzag periodic
nanotube in magnetic field, preprint arXiv:math.SP/0604007

\bibitem{korotyaev:06} E~Korotyaev and I.~Lobanov, Schr\"{o}dinger
  operators on zigzag nanotubes, Ann. Henri Poincaré 8 (2007), no. 6, 1151--1176.

\bibitem{KoS1}  V. Kostrykin and R. Schrader, Kirchhoff's rule for
quantum wires, J. Phys. A 32(1999), 595-630.

\bibitem{KoS2}  V. Kostrykin and R. Schrader, Kirchhoff's rule for
quantum wires. II: The inverse problem with possible applications to
quantum computers, Fortschr. Phys. 48(2000), 703--716.

\bibitem{KoS3}  V. Kostrykin and R. Schrader, The generalized star
product and the factorization of scattering matrices on graphs, J.
Math. Phys. 42(2001), 1563--1598.

\bibitem{KoSsemigr}V. Kostrykin, J.~Potthoff, and R.~Schrader, Contraction
Semigroups on Metric Graphs, in \cite{volume}.

\bibitem{KS}  T. Kottos and U. Smilansky, Quantum chaos on
graphs, Phys. Rev. Lett. 79(1997), 4794--4797.

\bibitem{KS2}  T. Kottos and U. Smilansky, Periodic orbit theory and
spectral statistics for quantum graphs, Ann. Phys. 274 (1999),
76--124.

\bibitem{KS3}  T. Kottos and U. Smilansky, Chaotic scattering on graphs,
Phys. Rev. Lett. 85(2000), 968--971.


\bibitem{Krejcirik} D.~Krej\v{c}i\v{r}\'{i}k, Twisting versus bending in
quantum waveguides, in \cite{volume}.

\bibitem{KrstVogtm} S.~Krsti\'{c}, K.~Vogtmann, Equivariant outer space and
automorphisms of free-by-finite groups, Comment. Math. Helv. {\bf 68}
(1993), no. 2, 216--262.

\bibitem{Ku89} P.~Kuchment, To the Floquet theory of periodic difference
equations, \textit{in Geometrical and Algebraical Aspects in
Several Complex Variables, Cetraro, Italy, 1989}, C. Berenstein
and D. Struppa (Editors), EditEl, 1991, 203--209.


\bibitem{Ku93}  P. Kuchment, \textit{Floquet Theory for Partial
Differential Equations}, Birkh\"{a}user, Basel 1993.

\bibitem{Ku01}  P. Kuchment, The Mathematics of Photonics Crystals, Ch.
7 in \textit{Mathematical Modeling in Optical Science}, Bao, G.,
Cowsar, L. and Masters, W.(Editors), 207--272, Philadelphia: SIAM,
2001.

\bibitem{Ku02}  P. Kuchment, Graph models of wave propagation in thin
structures, Waves in Random Media 12(2002), no. 4, R1-R24.

\bibitem{KuISAAC}  P. Kuchment, Differential and pseudo-differential operators
on graphs as models of mesoscopic systems, in Analysis and Applications,
H. Begehr, R. Gilbert, and M. W. Wang (Editors), Kluwer Acad. Publ. 2003, 7-30.

\bibitem{WPRM} P.~Kuchment (Editor), \textit{Waves in periodic and random media
(South Hadley, MA, 2002)}, Contemp. Math., v. 339, Amer. Math. Soc., Providence, RI, 2003.

\bibitem{WRM}  P. Kuchment (Editor), Quantum graphs and their applications, a special
issue of Waves in Random media, 14(2004), no.1.

\bibitem{Ku_graphs1} P.~Kuchment, Quantum Graphs I. Some basic structures, in \cite{WRM}, 2004,
S107--S128.

\bibitem{Ku_graphs2} P.~Kuchment, Quantum Graphs II. Some spectral properties
of quantum and combinatorial graphs, J. Phys. A 38 (2005), 4887-4900.

\bibitem{KK99}  P. Kuchment and L. Kunyansky, Spectral Properties of
High Contrast Band-Gap Materials and Operators on Graphs, Experimental
Mathematics, 8(1999), no.1, 1-28.

\bibitem{KK02}  P. Kuchment and L. Kunyansky, Differential operators on
graphs and photonic crystals, Adv. Comput. Math. 16(2002),
263-290.

\bibitem{KuchPin} P.~Kuchment and Y.~Pinchover, Liouville theorems and spectral
edge behavior on abelian coverings of compact manifolds, Trans. Amer. Math. Soc. {\bf 359}
(2007), no. 12, 5777--5815.


\bibitem{KuchPost} P.~Kuchment and O.~Post, On the spectra of carbon
nano-structures, Comm. Math. Phys. 275 (2007), no. 3, 805-826.

\bibitem{Ku_Vainb} P.~Kuchment and B.~Vainberg, On the structure of eigenfunctions
corresponding
to embedded eigenvalues of locally perturbed periodic graph
operators, Comm. Math. Phys. {\bf 268} (2006), 673-686.

\bibitem{KZ2} P.~Kuchment and H.~Zeng, Asymptotics of spectra of Neumann
Laplacians in thin domains, in Advances in Differential Equations
and Mathematical Physics, Yu. Karpeshina, G. Stolz, R. Weikard,
and Y. Zeng (Editors), Contemporary Mathematics v.387, AMS 2003,
199-213.

\bibitem{KurNow}P.~Kurasov and M.~Nowaczyk, Inverse spectral problem for quantum
graphs, J. Phys. A {\bf 38} (2005), no. 22, 4901--4915; Corrigendum: J. Phys.
A {\bf 39} (2006), no. 4, 993.

\bibitem{KuS}  P. Kurasov and F. Stenberg, On the inverse
scattering problem on branching graphs, J. Phys. A: Math. Gen.
35(2002), 101-121.

\bibitem{LLS} J.E. Lagnese, G. Leugering, E. Schmidt {\it Modelling, Analysis and
Control of
Multi-Link Flexible Structures}, Birkhauser, Boston, 1994

\bibitem{MNP}  V. Maz\'{y}a, S. Nazarov, and B. Plamenevskii, \textit{%
Asymptotic Theory of Elliptic Boundary Value Problems in
Singularly Perturbed Domains}, v. 1,2, Birkh\"{a}user Verlag,
2000. (English translation of the 1991 German edition by Akademie
Verlag)


\bibitem{AM1}  F. Ali Mehmeti, {\it A characterization of generalized
$C^{\infty}$ notion on nets,} Integral Eq. and Operator Theory, {\bf
9}, 1986, 753-766.

\bibitem{AM2}  F. Ali Mehmeti, {\it Regular solutions of transmission and
interaction problems for wave equations,} Math. Meth.  Appl. Sc.,
{\bf 11}, 1989, 665-685.



\bibitem{AM3}  F. Ali Mehmeti, {\it Nonlinear wave in networks,} Math. Res.
{\bf 80}, Akademie Verlag, 1994.

\bibitem{Mehmeti}F.~A.~Mehmeti, J.~von Below, and S.~Nicaise (Editors),
\textit{Partial differential equations on multistructures. Proceedings of the
International Conference held in Luminy, April 19--24, 1999}, Lecture Notes in
Pure and Applied Mathematics, 219. Marcel Dekker, Inc., New York, 2001.

\bibitem{MP}  Yu. B. Melnikov and B. S. Pavlov, Two-body scattering on a
graph and application to simple nanoelectronic devices, J. Math.
Phys. 36(1995), 2813-2825.

\bibitem{Pavl1} A.~B.~Mikhailova, B.~S.~Pavlov, and V.~I.~Ryzhii, Dirichlet-to-Neumann
techniques for the plasma-waves in a slot-diode, in \textit{Operator theory, analysis
and mathematical physics}, 77--103, Oper. Theory Adv. Appl., 174, Birkh\"{a}user, Basel, 2007.

\bibitem{ML}  R. Mittra, S. W. and Lee, \textit{Analytic Techniques in
the Theory of Guided Waves,} Collier - Macmillan, London 1971.

\bibitem{MW}  B. Mohar, W. Woess, A survey on spectra of infinite
graphs, Bull. London Math. Soc. 21 (1989), no. 3, 209--234.

\bibitem{MolchVainb_slow}S.~Molchanov and B.~Vainberg,  Slowing down of the wave
packets in quantum graphs, Waves Random Complex Media {\bf 15} (2005), no. 1, 101--112.

\bibitem{MolchVainb}S.~Molchanov and B.~Vainberg, Transition from a network of thin
fibers to the quantum graph: an explicitly solvable model, in \cite{BCFK}, 227--239, 2006.

\bibitem{Montroll}  E. Montroll, Quantum theory on a network, J. Math. Phys.
11 (1970), 635--648.

\bibitem{NS}  K.~Naimark and M.~Solomyak, Eigenvalue estimates for the
weighted Laplacian on metric trees, Proc. London Math. Soc.,
80(2000), 690--724.

\bibitem{NS2} K.~Naimark and M.~Solomyak, Geometry of Sobolev
spaces on the regular trees and hardy's inequalities, Russian J.
Math. Phys. {\bf 8} (2001), no. 3, 322-335.

\bibitem{Naim} M.~Naimark, \textit{Linear Differential Operators}, F. Ungar Pub.
Co., NY, 1967.

\bibitem{Nekr} V.~Nekrashevych and A.~Teplyaev, Groups and analysis on fractals,
in \cite{volume}.

\bibitem{Nic}  S. Nicaise, Some results on spectral theory over networks
applied to nerve impulse transmission, in \textit{Polynomes
Orthogonaux et Applicationes}, Lect. Not. Math. v. 1171,
Springer-Verlag, 1985, 532--541.

\bibitem{Nic2}  S. Nicaise and O. Penkin, Relationship between the lower
frequency spectrum of plates and networks of beams, Math. Methods
Appl. Sci. 23 (2000), no. 16, 1389--1399.

\bibitem{Nic4}  S. Nicaise and O.
Zair, Identifiability, stability and reconstruction results of point
sources by boundary measurements in heteregeneous trees, {\it
Revista Matematica de la Universidad Complutense de Madrid}, {\bf
16}, 2003, 1--28.

\bibitem{N1}  S.Novikov, Schr\"{o}dinger operators on graphs and
topology, Russian Math Surveys, 52(1997), no. 6, 177-178.

\bibitem{N2}  S.Novikov, Discrete Schr\"{o}dinger operators and
topology, Asian Math. J., 2(1999), no. 4, 841-853.

\bibitem{N3}  S.Novikov, Schr\"{o}dinger operators on graphs and
symplectic geometry, \textit{The Arnoldfest} \textit{(Toronto, ON, 1997)},
397--413, Fields Inst. Commun., 24, Amer. Math. Soc., Providence,
RI, 1999.

\bibitem{Paiv} P.~Ola and L.~Paivarinta, Mellin operators and pseudodifferential
operators on graphs, in \cite{WRM},
S129--S142, 2004.

\bibitem{Pavl2} V. L. Oleinik, B. S. Pavlov, and N. V. Sibirev,  Analysis of the
dispersion equation for the Schr\"{o}dinger operator on periodic metric graphs,
in \cite{WRM}, 157--183, 2004.

\bibitem{Ong} B.~Ong, On the limiting absorption principle and
spectra of quantum graphs, in \cite{BCFK}, 241--249.

\bibitem{Pankr} K.~Pankrashkin, Spectra of Schr\"{o}dinger operators on
equilateral quantum graphs, Lett. Math. Phys. 77 (2006) 139-154.

\bibitem{Pankr2} K.~Pankrashkin, Localization effects in a periodic quantum
graph with magnetic field and spin-orbit interaction,
J. Math. Phys. {\bf 47} (2006), 112105.

\bibitem{Pav2}  B. S. Pavlov, The theory of extensions and explicitly
solvable models, Russian. Math. Surveys 42(1987), 127-168.

\bibitem{PF}  B. S. Pavlov and M. D. Faddeev, A model of free electrons
and the scattering problem, Theor. and Math. Phys. 55(1983), no.2,
485--492.

\bibitem{Pavl4}  B.~Pavlov Resonance quantum switch: matching domains, in
\textit{Surveys in analysis and operator theory (Canberra, 2001)}, 127--156,
Proc. Centre Math. Appl. Austral. Nat. Univ., 40, Austral. Nat. Univ.,
Canberra, 2002.

\bibitem{Pavl3}  B.~Pavlov and K.~Robert, Resonance optical switch: calculation
of resonance eigenvalues, in \cite{WPRM} 141--169, 2003.

\bibitem{Penkin} O.~M.~Penkin, Second-order elliptic equations on a stratified set.
Differential equations on networks, J. Math. Sci. (N. Y.) 119 (2004), no. 6, 836--867.

\bibitem{Piv}  V. Pivovarchik, Inverse problem for the Sturm-Liouville
equation on a simple graph, SIAM J. Math. Anal. 32 (2000), no. 4,
801--819

\bibitem{Pokornyi} Yu.~V.~Pokornyi et al, \textit{Differential equations on geometrical
graphs}, Moscow: Fizmatlit, 2004. (in Russian)

\bibitem{Post} O. Post, Equilateral quantum graphs and boundary triples, in \cite{volume}.

\bibitem{Rab}V.~Rabinovich, S.~Roch, and B.~Silbermann, \textit{Limit Operators
and Their Applications in Operator Theory}, Operator Theory: Advances and Applications,
150. Birkh\"{a}user Verlag, Basel, 2004.

\bibitem{RS}  M. Reed and B. Simon, \textit{Methods of Modern
Mathematical Physics} v. 4, Acad. Press, NY 1978.

\bibitem{RiB}  M. J. Richardson and N. L. Balazs, On the network model
of molecules and solids, Ann. Phys. 73(1972), 308-325.


\bibitem{Roth}  J.-P. Roth, Le spectre du laplacien sur un graphe,
Th\'{e}orie du potentiel (Orsay, 1983), Lecture Notes in Math., v.
1096, Springer-Verlag, Berlin, 1984, 521--539.


\bibitem{RS2}  J. Rubinstein and M. Schatzman, Asymptotics for thin
superconducting rings, J. Math. Pures Appl. (9) 77 (1998), no. 8,
801--820.

\bibitem{RS3}  J. Rubinstein and M. Schatzman, On multiply connected
mesoscopic superconducting structures, S\'{e}min. Th\'{e}or.
Spectr. G\'{e}om., no. 15, Univ. Grenoble I,
Saint-Martin-d'H\`{e}res, 1998, 207-220.

\bibitem{RS4}  J. Rubinstein and M. Schatzman, Variational problems on
multiply connected thin strips. I. Basic estimates and convergence
of the Laplacian spectrum. Arch. Ration. Mech. Anal. 160 (2001),
no. 4, 271--308.

\bibitem{RuS}  K. Ruedenberg and C. W. Scherr, Free-electron network
model for conjugated systems. I. Theory, J. Chem. Physics, 21(1953),
no.9, 1565-1581.

\bibitem{Schap} P.~Schapotschnikow and S.~Gnutzmann, Spectra of graphs and semi-conducting
polymers, in \cite{volume}.

\bibitem{AiS}  J. Schenker and M. Aizenman, The creation of spectral
gaps by graph decoration, Lett. Math. Phys. 53 (2000), no. 3, 253.

\bibitem{Schnol} E.~Schnol, On the behavior of Schr\"{o}dinger
equation, Mat. Sbornik {\bf 42} (1957), 273--286 (in Russian).


\bibitem{Shubin_pseudodiff} M.~A.~Shubin, Pseudodifference operators and
their Green's functions, Math. USSR Izvestiya {\bf 26} (1986), no.
1, 605--622.

\bibitem{BerezShub} M.~A.~Shubin and F.~A.~Berezin, \textit{The Scr\"{o}dinger Equation},
(Mathematics and Its Applications, Soviet Series, Vol 66), Kluwer
2002.

\bibitem{Solom03} M.~Solomyak,  Laplace and Schr\"{o}dinger operators on
regular metric trees: the discrete spectrum case, in  \textit{Function spaces,
differential operators and nonlinear analysis (Teistungen, 2001)}, 161--181,
Birkh\"{a}user, Basel, 2003.

\bibitem{Solom} M.~Solomyak, On the spectrum of the Laplacian on regular metric
trees, in \cite{WRM}, S155--S171, 2004.

\bibitem{SoSo}  A. V. Sobolev and M. Solomyak, Schr\"{o}dinger operator
on homogeneous metric trees: spectrum in gaps, Rev. Math. Phys.
{\bf 14} (2002)421--467.

\bibitem{Strichartz} R.~Strichartz, \textit{Differential Equations on Fractals: A
Tutorial}, Princeton University Press 2006.

\bibitem{Sunada_book} T.~Sunada, \textit{Why do diamonds look so beautiful? -
Introduction to discrete harmonic analysis}, Springer Japan 2006.

\bibitem{Sunada} T.~Sunada, Discrete Geometric Analysis, in \cite{volume}.

\bibitem{Thor}  T. J. Thornton, Mesoscopic devices, Rep. Prog. Phys.
57(1994), 311-364.

\bibitem{Vogt} K. Vogtmann, Automorphisms of free groups and outer space,
Proc. of Conf. on
Geometric and Combinatorial Group Theory, Part I (Haifa, 2000),
Geom. Dedicata 94 (2002), 1--31.

\bibitem{wilson} J.~H.~Wilson, \textit{Vacuum Energy in Quantum Graphs},
Undergraduate Research Fellow thesis, Texas A\& M University, 2007.
 http://handle.tamu.edu/1969.1/5682

\bibitem{Winn} B.~Winn, A conditionally convergent trace formula for quantum
graphs, in \cite{volume}.


\bibitem{X}  J.-B.Xia, Quantum waveguide theory for mesoscopic
structures, Phys. Rev. B 45(1992), 3593-3599.


\bibitem{Yurko} V.\,Yurko, Inverse spectral problems for
Sturm--Lioville operators on graphs, {\it Inverse Problems}, {\bf
21} (2005) 1075--1086.

\end{thebibliography}
\end{document}